\documentclass[amsmath,amssymb,aps,twocolumn,notitlepage,showpacs,pra,floatfix,longbibliography]{revtex4-1}
\usepackage{graphicx}
\usepackage{bm}
\usepackage{hyperref}
\usepackage[usenames, dvipsnames]{color}
\usepackage{verbatim}

\hypersetup{
	colorlinks=true,
	linkcolor=blue,
	citecolor=blue,
	filecolor=blue,
	urlcolor=blue,
}

\begin{document}
	
	\title{Anisotropic long-range interaction investigated with cold atoms}

	\author{Vincent Mancois$^{2,3}$, Julien Barr\'e$^{4}$,  Chang Chi Kwong$^{2}$, Alain Olivetti$^{6}$, Pascal Viot$^{3,1}$, and David Wilkowski$^{1,2,5}$}
	\affiliation{$^1$ MajuLab, International Joint Research Unit UMI 3654, CNRS, Universit\'e C\^ote d'Azur, Sorbonne Universit\'e, National University of Singapore, Nanyang Technological University, Singapore}
	\affiliation{$^2$ PAP, School of   Physical and Mathematical Sciences, Nanyang Technological University, 637371 Singapore}
	\affiliation{$^3$ Laboratoire de Physique
		Th\'eorique de la Mati\`ere Condens\'ee, Sorbonne Universit\'e, CNRS  UMR 7600,  4, place Jussieu, 75005 Paris, France}
	\affiliation{$^4$ Institut Denis Poisson, Universit\'e d'Orl\'eans, CNRS, Universit\'e de Tours, et Institut Universitaire de France}
	\affiliation{$^5$ Centre for Quantum Technologies, National University of Singapore, 117543 Singapore}
	\affiliation{$^6$ Universit\'e C\^ote d'Azur, CNRS, LJAD, 06108 Nice, France}
	
	\begin{abstract}

		In two dimensions, a system of self-gravitating particles collapses and forms a singularity in finite time below a critical temperature $T_c$. We investigate experimentally 
		a quasi two-dimensional cloud of cold neutral atoms in interaction with two pairs of perpendicular counter-propagating quasi-resonant laser beams, in order to look for
		 a signature of this ideal phase transition: indeed, the radiation pressure forces exerted by the laser beams can be viewed as an anisotropic, 
		 and non-potential, generalization of two-dimensional self-gravity. We first show that our experiment operates in a parameter range which should be
		  suitable to observe the collapse transition. 
		However, the experiment unveils only  a moderate compression instead of a phase transition between the two phases.
		A three-dimensional numerical simulation shows that both the finite small thickness of the cloud, which induces a competition between the effective gravity force and the repulsive force due to multiple scattering, and the atomic losses due to heating in the third dimension, contribute to smearing the transition.
	\end{abstract}
	\date{\today}
	\maketitle
	
	\section{Introduction}
	When particles interact with a force decaying at large distance like $r^{-\alpha}$ where $\alpha$ is less than the space dimension, the force is long-range, and the system displays some intriguing features 
	both at and out of equilibrium \cite{Campa2014}. However, these systems, especially those involving attractive forces, are often not easily accessible experimentally. 
	
	Since near resonant laser beams induce
	effective long-range interactions in cold atomic clouds \cite{Dalibard1985,Sesko1991}, it has been suggested that they could be original experimental testbeds for long-range interactions. There are two types of effective long-range forces. First, Dalibard \cite{Dalibard1988} identified the so-called "shadow effect" in cold atomic clouds trapped by counter-propagating laser beams. Here, absorption of the near resonant laser beams, as they propagate
	inside the cloud, creates an intensity imbalance between the two counter-propagating beams, resulting in an effective long-range attraction between atoms. In the small optical depth regime, this force is similar to one dimensional (1D) gravity, \textit{i.e.} the force between two atoms does not depend on the distance. In standard three dimensional (3D) optical molasses, there are three orthogonal pairs of counter-propagating beams, and the combined shadow effect looks like three 1D gravitational interactions directed along each pair of beams. In particular, although the force is anisotropic and does not derive from a potential, its divergence is identical to gravity: it may then trigger a "pseudo gravitational collapse" \cite{Dalibard1988}. However, a few years after the shadow effect was identified,
	D. Sesko \textit{et al}.  \cite{Sesko1991} showed that multiple scattering of photons inside the clouds also creates an effective long-range force, but of repulsive nature, similar to a Coulomb force in the small optical depth regime. 
	This repulsive interaction is typically of the same order of magnitude 
	as the shadow effect, but generally slightly stronger. Then, the shadow effect merely renormalizes
	 the repulsive force, and  
	its exotic signatures are difficult to pinpoint. As the repulsive force typically dominates,
	 the cloud rather behaves as a non-neutral plasma \cite{Mendonca2008,Romain2011,Tercas2013}.
	
	Adding anisotropic traps, one can modify the geometry of the cloud in order to decrease the strength of 
	the repulsive force, and ultimately make the shadow
	effect dominant. For instance, Chalony {\it et al.} \cite{Chalony2013} have  argued theoretically and demonstrated experimentally that laser induced interactions 
	in a thin cigar-shaped cloud bear similarity with 1D gravity. Similarly, 
	 Barr\'e {\it et al.} \cite{Barre2014} have suggested that the shadow effect could be dominant in a 
	thin pancake-shaped cloud and, neglecting multiple scattering and the resulting repulsive force, 
		argue that a collapse transition may occur if the attractive force is strong enough. 
	Since the divergence of the attractive force is identical to the gravity case, 
	such a collapse would be similar to the one happening for $2D$ self-gravitating systems 
	in the canonical ensemble \cite{Chavanis2004}, or in the Keller-Segel model of bacterial 
	chemotaxis \cite{Keller1970,Keller1971}, but with a non-potential force.
	
	We report here on an experiment inspired by Ref. \cite{Barre2014}: a cold atomic 
	cloud is loaded in a very flat, pancake-shaped, optical trap, \textit{i.e.} with one very stiff direction.
	 It is then subjected to two perpendicular pairs of counter-propagating laser beams in the easy plane of
	  the trap. We observe a fast but moderate compression of the cloud, whereas a $2D$ model 
	  predicts 
	  further compression eventually leading to a collapse of the atomic cloud. 
	  A more realistic description of the experiment is proposed by introducing a $3D$ model which includes 
	  the finite thickness of the cold atom,  the associated repulsive forces due to the multiple scattering 
	  and some atoms losses due to the finite depth of the optical dipole trap.
	
	The paper is organized as follows. Since the shadow effect in a pancake-shaped cloud bears similarity with
	 2D gravity, we review this analogy in Sec. \ref{sec:sga}. We first present the $2D$ model of self-gravitating
	  particles, and its phase diagram in the canonical ensemble, which shows a collapse transition 
	  below a critical temperature. We compare it with the $2D$ model of the shadow effect,	
	  where the attractive force does not derive from a potential, as opposed to its true $2D$ 
	  gravity counterpart. We then identify the parameters for which a collapse should be observed, 
	  based on the simplified $2D$ analysis.
	In Sec. \ref{sec:experiment}, we present the experimental set-up and the results showing a finite 
	compression of the cloud due to the attractive force, but not as strong as 
	picted in Sec. \ref{sec:sga}. We highlight the presence of atoms losses due to spontaneous 
	emission heating in the third direction.	
	In Sec. \ref{sec:model}, we attempt to bridge the gap between the simplified $2D$ model of 
	Sec. \ref{sec:sga} and the actual experiments of Sec. \ref{sec:experiment}, 
	by considering more realistic 3D models. The first model takes into account the finite thickness of 
	the cold atomic cloud in the third dimension and the repulsive Coulomb-like force. 
	We numerically solve the associated Smoluchowski-Poisson equation, and analyze how the $2D$ collapse 
	transition is smeared out by this finite thickness of the cloud. We  also propose a 
	phenomenological extension of the previous 3D model accounting for the heating due to spontaneous emission, 
	which provides an estimate of the typical time to spill the atoms out of the external optical trap in 
	the third direction, orthogonal to the pancake-shaped trap. 
	This more realistic model qualitatively reproduces the experimental results. 
	In the conclusion, we suggest some possible improvements on the experimental set-up in order to 
	reach larger compression.

	\section{Two-dimensional models: self-gravitating systems, chemotaxis and cold atoms}\label{sec:sga}
	In order to highlight  the universality of the collapse phenomenon for systems interacting 
	with gravitational (or quasi-gravitational) forces, we first briefly review the well-studied 
	self-gravitating and chemotaxis systems before introducing the cold atom system for which a 
	similar behavior is expected.
	\subsection{$2D$ self-gravitating systems}
	\label{sec:2Dsgs}

	For a two-dimensional system of thermalized, self-gravitating Brownian 
		particles, the dynamics is described by the overdamped limit of the Langevin 
		equations (see for instance \cite{Chavanis2007a})
	\begin{equation}\label{eq:dynam}
	\dot{\bm{r}_i}=\frac{Gm}{\eta}\sum_{j\neq i} \frac{\bm{r}_j-\bm{r}_i}{|\bm{r}_j-\bm{r}_i|^2} +\sqrt{\frac{2 k_B \theta}{m \eta}} \chi_i(t),
	\end{equation}
	where $\bm{r}_i$ and $m$ are the  position and mass of particle $i$, $G$ is the gravitational coupling,  $\eta$ is the viscous friction coefficient, $\theta$ the temperature, and $k_B$ the Boltzmann constant. The $\chi_i$ are independent Gaussian white 
	noises satisfying $\langle\chi_i(t)\rangle=0$ and $\langle\chi_i(t)\chi_j(t')\rangle=\delta_{ij}\delta(t-t')$.
	
	We introduce dimensionless variables $\bm{x}_i=\bm{r}_i/L_i,~s= t/\tau$, and $\xi_i(s)=\tau^{1/2}\chi_i(t\equiv \tau s) $. 
	Then $\langle\xi_i(s)\rangle=0$ and $\langle\xi_i(s)\xi_j(s')\rangle=\delta_{ij}\delta(s-s')$, 
	and Eq.~(\ref{eq:dynam}) becomes
	\begin{equation}
	\frac{d\bm{x}_i}{ds} = \frac{2}{\pi}\sum_{i\neq j} \frac{\bm{x}_j-\bm{x}_i}{|\bm{x}_j-\bm{x}_i|^2} +\sqrt{2T}
	 \xi_i(s)   \label{eq:2Dgrav}\\
	\end{equation}
	with
	\begin{equation}
	\frac{\tau}{L^2}=\frac{2}{\pi}\frac{\eta}{Gm},~T=\frac{2k_B\theta}{\pi Gm^2}.
	\end{equation}
	Thus, the motion is controlled by a single parameter, the dimensionless temperature $T$. We note also that only the ratio $\tau/L^2$ appears, so there is still some freedom in the choice of $\tau$ and $L$ which will be used later
	 to rewrite the external harmonic trap confinement in a dimensionless way [see Eq. \eqref{eq:2Drho}]. 
	 The unusual $2/\pi$ factor is introduced to facilitate the comparison with the shadow effect
	  in cold atomic clouds (see Sec. \ref{sec:sgb} below). 
	
	Associated with the Langevin description  of the system [see Eq. \eqref{eq:2Dgrav}], 
	the time evolution of the density $\rho_{\rm 2D}({ \bm{x},s})$ is governed at the mean-field level
	 by a Smoluchowski-Poisson equation
	
	\begin{equation}\label{eq:SP}
	\frac{\partial \rho_{\rm 2D}}{\partial s}=\bm{\nabla} \cdot \left[\rho_{\rm 2D}\bm{\nabla}\Phi+T \bm{\nabla}\rho_{\rm 2D}\right],
	\end{equation} 
	where $\Phi$ is the mean gravitational potential induced by the particles that satisfies the Poisson equation
	\begin{equation}\label{eq:Poisson}
	\Delta \Phi =4 \rho_{\rm 2D}.	
	\end{equation}
	It turns out that Eq. (\ref{eq:SP}) has a critical temperature $T_c=1/(2\pi)$ \cite{Chavanis2004}.
	 For $T>T_c$, the system receives heat from the thermostat, which is transformed continuously in 
	 potential energy, and the system expands continuously without limit \cite{Chavanis2006,Chavanis2006b}. 
	 In a bounded domain, or in the presence of a confining potential (absent in Eq. (\ref{eq:SP})),
	  a stable equilibrium is eventually reached. Conversely for $T<T_c$, the heat flows from the system to the
	   reservoir, the system 
	shrinks and develops a singularity in finite time. 
	This low temperature phase can be stabilized by a short-range repulsive interaction: 
	the collapse transition is then replaced by a transition with a formation of a dense core \cite{Chavanis2014}. 
	
	\subsection{Chemotaxis}
	In biology, interaction between organisms (bacteria, amoebae, cells) may be driven by chemotaxis \cite{Keller1970,Keller1971,Chavanis2007,Chavanis2010b,Dyachenko2013,Raphaeel2014}. A simple stochastic version of these models is described by a Smoluchowski equation for the density of organisms $\rho_\text{2D}$ as in Eq. (\ref{eq:SP}), where the mean potential is replaced by (minus) a density of secreted chemical $n({\bf r},t)$. The time evolution of the chemical is given by a reaction-diffusion equation.
	\begin{equation}\label{eq:rde}
	\frac{\partial n}{\partial t}=D_c\Delta n -kn+\lambda \rho_{\rm 2D},
	\end{equation}
	where $D_c$ is the diffusion constant of the chemical, $k$ its rate of degradation and $\lambda$ its rate of production.
	If the dynamics of the chemical  is fast with respect to the dynamics of the density $\rho_{\rm 2D}$, and if the degradation rate can be neglected, 
	Eq. \eqref{eq:rde} can be replaced by a Poisson equation similar to Eq. (\ref{eq:Poisson}), and the dynamics of the bacterial density is described by the same system as the 2D self-gravitating particles.

	\subsection{Cold atoms as a quasi 2D self-gravitating system}\label{sec:sgb}
	
\begin{figure}[t!]
		\includegraphics[width=0.46\textwidth]{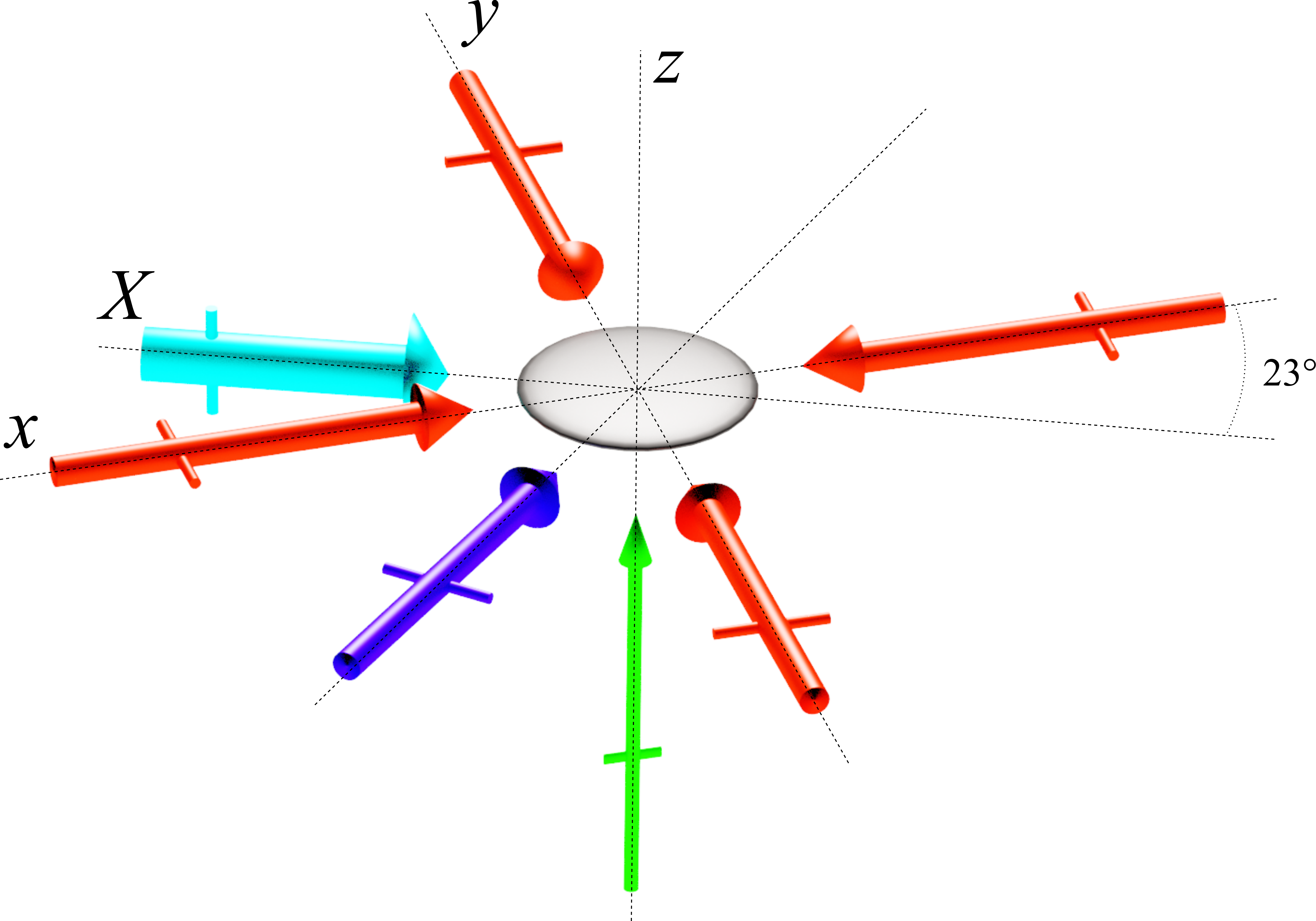}
		\caption{Depiction of the beam configuration. The atomic cloud (grey) is confined in a horizontal pancake-shaped optical trap, obtained with a focused elliptical far-off-resonance beam (cyan arrow). The 2D artificial "pseudo-gravity" is created using the four contra-propagating LRI beams (red arrows). The cloud is imaged with a resonant probe (blue). All those beams are propagating in the horizontal plane. Another dipole beam (green arrow), propagating along the vertical axis, increases the cloud's initial density. The linear polarization axis of each beam is indicated by a rigid bar normal to the arrows.}
		\label{fig:setup}
	\end{figure}

	We now consider a cold atom cloud confined in a pancake-shaped strongly anisotropic harmonic trap, see Fig.~\ref{fig:setup}. The vertical size $L_z$ is supposed to be so small that the system can be reduced to a quasi 2D system in the $xy$ plane. Two orthogonal pairs of counter-propagating laser beams, near-resonance but red detuned with respect to an atomic transition, form  a 2D optical molasses in the $xy$ plane: We shall refer to them as the long range interaction (LRI) beams. For such a geometry, it is suggested in \cite{Barre2014} that the interactions inside the cloud are dominated by the shadow effect. Indeed, the vertical dimension offers a route for scattered photons to escape the cloud before being reabsorbed. This effect is reinforced choosing the laser polarization within the $xy$ plane. Thus, our model neglects multiple scattering and its associated repulsive force. Then the attractive shadow effect, albeit being non conservative, bears similarities with gravity. The system may exhibit an extended/collapsed phase transition at a critical temperature, similar to the thermalized self-gravitating systems and chemotactic models. 
	
	The near-resonant lasers are along the $\hat{x}$ and $\hat{y}$ axes with the same optical frequency
	 $\omega_L$. Their interaction with atoms is characterized by an on-resonance saturation parameter 
	 $s_0=I/I_s$, with $I$ the laser intensity per beam and $I_s$ the saturation intensity. They address a closed two-level transition of linewidth $\Gamma$ with a detuning $\bar{\delta}=(\omega_L-\omega_0)/\Gamma<0$. The radiation pressure component of the light-atom interaction for a low optical depth and a low saturation parameter can be written \cite{Chalony2013}
	
	\begin{equation}
	\mathbf{F}_\text{3D}(\mathbf{r})=\hbar k \frac{\Gamma}{2}s_0 
	\left( \frac{[ \overset{\leftharpoonup}{b_x}(\mathbf{r})-\overset{\rightharpoonup}{b_x}(\mathbf{r}) ] 
		\hat{x} + [ \overset{\leftharpoonup}{b_y}(\mathbf{r}) 
		-\overset{\rightharpoonup}{b_y}(\mathbf{r})] \hat{y} }{1+4\bar{\delta}^2}\right),
	\label{eq:gravitational_force_2D}
	\end{equation}
	where the optical depth at position  $\mathbf{r}=(x,y,z)$ seen by the laser coming from $-\infty$ is given by
	\begin{equation}\label{eq:opticalt}
	\overset{\rightharpoonup}{b_x}(\mathbf{r})=\frac{\sigma_0 N}{(1+4\bar{\delta}^2)}
	\int_{-\infty}^x\rm{d}x'\rho(\mathbf{r'}),
	\end{equation} 
	with $\sigma_0=\frac{6\pi}{k^2}$ the on-resonance absorption cross-section, $k$ the wavenumber, and $N$ the atom number.  
	$\overset{\leftharpoonup}{b_x}(\mathbf{r})$, corresponding to the contra-propagating laser beam, is obtained modifying the integration range in
	Eq. \eqref{eq:opticalt} 
	to $[x, +\infty)$. 
	$\overset{\rightharpoonup}{b_y}(\mathbf{r})$ and $\overset{\leftharpoonup}{b_y}(\mathbf{r})$
	have similar definitions, swapping the role of $x$ and $y$. 
	We assume the equilibrium in the transverse direction to be reached quickly, 
	so the normalized density is written $\rho(x,y,z,t) = \rho_{\rm 2D}(x,y,t) 
	(2\pi L_z^2)^{-1/2}e^{-\frac{z^2}{2L_z^2}}$, where $L_z$ is the transverse size of the cloud, 
	assumed to be small and constant. 
	Inserting into Eqs. \eqref{eq:gravitational_force_2D} and \eqref{eq:opticalt}, 
	and averaging over the transverse direction with weight $(2\pi L_z^2)^{-1/2}e^{-\frac{z^2}{2 L_z^2}}$, 
	one obtains an expression for the 
	effective force in $2D$:
	\begin{align} 
	\label{eq:F_2D}
	\mathbf{F}_\text{2D}(\bm{r})&=\left( \begin{array}{c} -C\int {\rm sgn}(x-x')\rho_{\rm 2D}(x',y)dx' \\
	 -C\int {\rm sgn}(y-y') \rho_{\rm 2D}(x,y')dy' \end{array} \right),
	\end{align}
	with 
	\begin{equation}	\label{eq:C}
	C=\frac{\hbar k \Gamma}{2}s_0 \frac{N}{ 2\sqrt{\pi}L_z}\frac{\sigma_0}{(1+4\bar{\delta}^2)^2}.
	\end{equation}
	
	The friction force due to Doppler cooling associated with the LRI beams is given by $\mathbf{F}_\text{d}=-m\eta \mathbf{v}$ 
	where the friction coefficient $\eta$ is given as in Ref. \cite{Lett_1989} in the low saturation limit by
	\begin{equation}
	\eta = -8 s_0 \frac{\hbar k^2}{m}  \frac{\bar{\delta}}{(1+4\bar{\delta}^2)}.
	\end{equation}
	The friction is typically strong enough to warrant an overdamped description of the atomic 
	cloud (see Ref.\cite{Lett_1989} for a detailed discussion).

	We introduce the two parameters $\tau=\eta/\omega^2$ and $L=\sqrt{C/(m\omega^2)}$, 
	with $\omega$ the in-plane harmonic trap frequency in the $xy$ plane, and the dimensionless 
	variables: $s=t/\tau$, $\mathbf{r'}=\mathbf{r}/L$, $\rho_{\rm 2D}'=\rho_{\rm 2D} L^2$, 
	and $\mathbf{F}'_{2D}=\mathbf{F}_{2D}/m\omega^2$. 
	Then, taking into account the 2D approximation of the force due to the shadow
		 effect
		 Eq.\eqref{eq:F_2D}, the trapping force and the temperature, the overdamped equation of motion 
		 can be expressed as a continuity equation (see \cite{Klimontovich1994} 
		 for a detailed derivation):
	
	\begin{eqnarray} 
	\frac{\partial \rho_{\rm 2D}'}{\partial s} &=& \bm{\nabla'} \cdot \left[ -\rho_{\rm 2D}' (\mathbf{F}'_{2D}(\bm{r'})- \bm{r'}) + T \bm{\nabla'} \rho_{\rm 2D}' \right], \label{eq:2Drho}
	\end{eqnarray}
	where $T=\frac{k_B\Theta}{C} $ is the dimensionless temperature, and we get
	\begin{equation}	\label{eq:2DF}
	\bm{\nabla'} \cdot \mathbf{F}'_{2D}(\bm{r'})=-4\rho_{\rm 2D}'.
	\end{equation}		 
	We note that the system of Eqs. \eqref{eq:2Drho}-\eqref{eq:2DF} has a form similar to
	 the Smoluchowski-Poisson system of Eqs. \eqref{eq:SP}-\eqref{eq:Poisson}. Indeed, the divergence of the 
	 interaction force is the same in both cases,
	  but the long-range force is now anisotropic and does not derive from a potential.
	   Since Eq.\eqref{eq:F_2D} has the form of a 1D gravitational interaction along each axis,
	   	 we shall refer to 
	   	Eqs.\eqref{eq:2Drho}-\eqref{eq:2DF} as the "1D+1D" gravitational model.
   	 The additional harmonic force ensures the stability of the high-temperature phase, as already mentioned
   	  in the previous section.
	
	It is shown in Ref. \cite{Barre2014,Barre2019} that the system of Eqs. \eqref{eq:2Drho}-\eqref{eq:2DF} 
	is stable at high temperatures: the diffusion wins over the attraction and equilibrates 
	the external harmonic confinement. It is also suggested in \cite{Barre2014} that
	the system undergoes a collapse transition in a finite time for $T<T_c$, where $T_c$ is the critical 
	temperature of the transition. 
	Numerical simulations allow us to estimate the transition temperature and 
	give $T_c \approx 0.13-0.15$, to be compared with the 2D gravitational model 
	for which $T_c=1/(2\pi)\simeq 0.159 $. The critical temperature for
	 the "1D+1D" gravitational model Eqs. \eqref{eq:2Drho}-\eqref{eq:2DF} seems to be slightly lower
	  than for the standard 2D gravitational model 
	  Eqs.\eqref{eq:SP}-\eqref{eq:Poisson}.
	
	\begin{figure}[t!]
		\centering
		\includegraphics[width=0.44\textwidth]{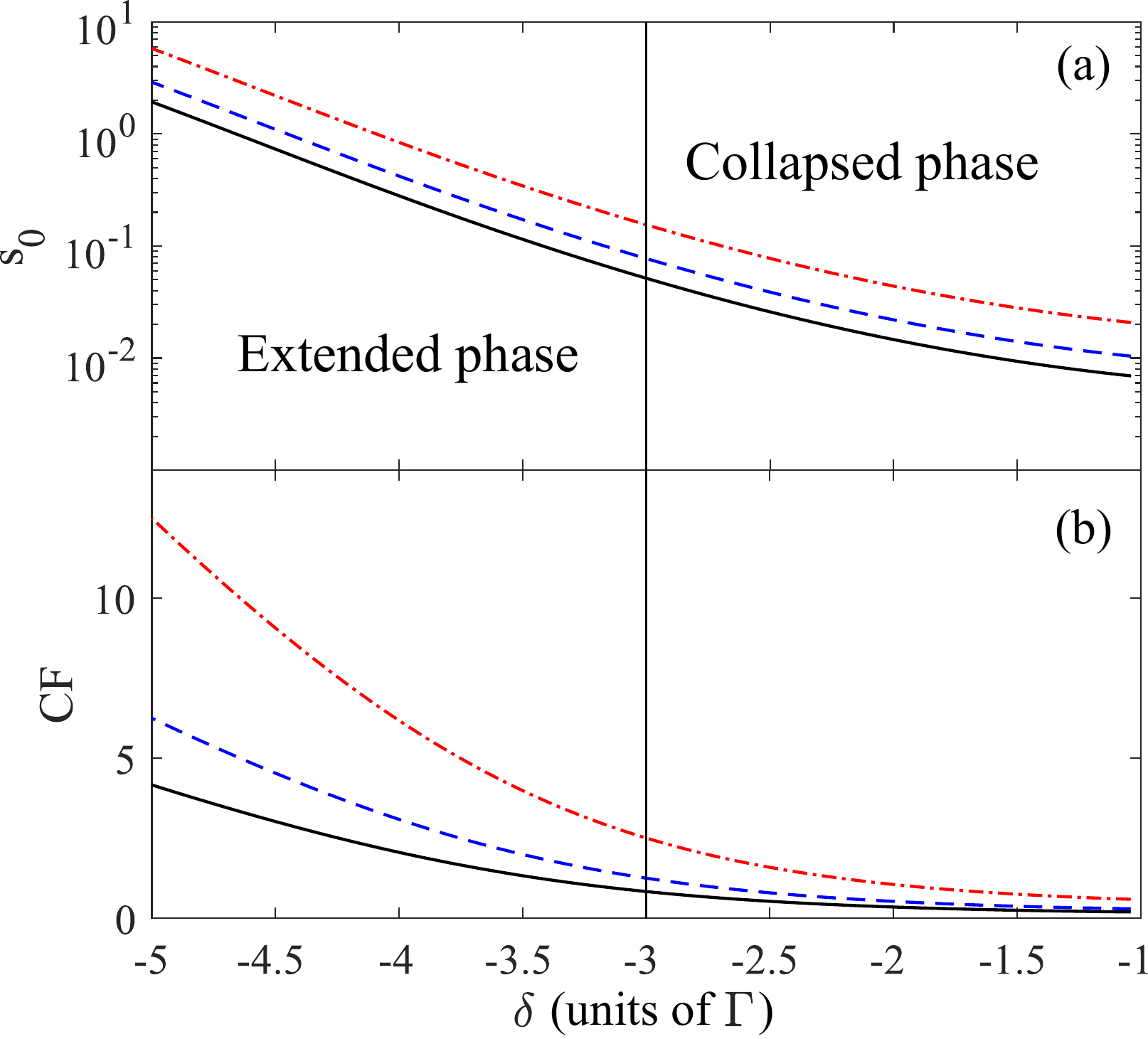}
		\caption[$~~$2D Phase diagram]{(a) Phase diagram of the 2D collapse transition. 
			Above the curves, the model predicts a collapsed phase whereas the extended phase lies below.
			 We use Eq. \eqref{eq:C} with a critical dimensionless temperature of $T_c=0.14$ 
			 and a temperature $\theta=1\,\mu$K. (b) The expected compression factor CF in the collapsed phase 
			 (see text for more details). The full, dashed and dotted-dashed curves 
			 correspond to an atoms number of $N=(3,\,2,\,1)\times 10^5$, respectively. 
			 Parameters of Eq.\eqref{eq:compf} are chosen from the experimental setup (see text). 
			 The vertical black line indicates $\bar{\delta}=-3$, the detuning used in the experiment. }
		\label{fig:phase_diagram_2D}
	\end{figure}
	
	Figure \ref{fig:phase_diagram_2D} (a) shows the phase diagram of the 2D model, where the plain black, dashed blue, and dotted-dashed red curves correspond to critical lines for atoms number of $N=(3,\,2,\,1)\times 10^5$ respectively. Parameters are chosen according to the experiment: the gas temperature is $\theta=1\,\mu$K and the trapping frequencies are $\omega=20\,$Hz, and $\omega_z=300\,$Hz in the horizontal plane, and along the vertical axis, respectively. The critical lines are obtained setting a dimensionless critical temperature $T_c=0.14$. Above the critical line the system is expected to be in the collapsed phase, and in the extended phase below it. The vertical line corresponds to a laser frequency detuning of $\bar{\delta}=-3$ as for the experiment (see Sec. \ref{sec:experiment}). At this detuning, we observe that the phase transition is predicted for a saturation parameter of $s_0\simeq 0.2$, the exact value depending on the atoms number. Those parameters are easily achieved in the experiment.
	
	We now discuss what should be the experimental signature of this collapsed regime.	
	The expression Eq. \eqref{eq:2DF} for the attractive long-range interaction force relies on 
	the linearization of the laser absorption. 
	In particular, it is not valid anymore as soon as
	the optical depth reaches values of order one, which will happen as the cloud contracts. 
	In this case, the shadow effect becomes weaker near the center of the cloud, and not long-range anymore; 
	we then expect a finite compression of the cloud with an optical depth saturated to a value 
	of order one. Setting the peak optical depth to one provides an
	 order of magnitude
	of the cloud's size in the collapsed phase
	\begin{equation}\label{eq:compf}
	L=\frac{N}{2 \pi L_z}\frac{\sigma_0}{(1+4\bar{\delta}^2)}.
	\end{equation}
	For the sake of simplicity, we have considered here an isotropic cloud in the $xy$ plane with
	 Gaussian profiles. We define a compression factor as 
	\begin{equation}\label{eq:CF}
	\text{CF}=L^{\rm th}/L,
	\end{equation}
	where $L^{\rm th}=\omega^{-1}\sqrt{k_B\theta/m}$ is the cloud's size in a harmonic trap of 
	frequency $\omega$ at thermal equilibrium, without long-range force. 
	Expected compression factors in the collapsed phase correspond to the curves in Fig. \ref{fig:phase_diagram_2D}b. Importantly, the compression factor increases as the atoms number decrease. Thus, in the experiment, where atoms losses are present (see Sec. \ref{sec:evaporation}), the collapsed phase is expected to be characterized by an increasing compression factor together with a saturated optical depth, as time increases (and atoms are lost). When too many atoms are lost, the system should finally leave the collapsed phase and its size increases.
	
	We note in Figure \ref{fig:phase_diagram_2D} (b), that CF can be smaller than one (see region where $|\bar{\delta}|$ is small). It means that the cloud in the harmonic trap has an optical depth larger than one without long-range force. In this situation, the LRI lasers are not expected to play a significant role. Therefore, in this region, Eq.\eqref{eq:compf} is no longer valid and CF must be replaced by one.
	
	\section{Experiment}
	\label{sec:experiment}
	\subsection{Cloud preparation}

	The atomic system consists in a laser cooled atomic cloud of $^{88}$Sr \cite{Loftus2004,Mickelson2009,Shao-Kai2009,Chalony2011}. The detailed two-stages cooling in a magneto-optical trap (MOT) is presented in Ref. \cite{Yang2015}. The last stage of the cooling scheme as well as the 2D artificial gravity are obtained with red detuned lasers addressing the $^1$S$_0$ $\rightarrow$ $^3$P$_1$ intercombination line of natural linewidth $\Gamma=2\pi \times 7.5$ kHz at $\lambda=689$ nm.
	
	After the final cooling stage, atoms are transferred into a single beam horizontal optical dipole trap (ODT) at 925 nm linearly polarized along $\hat{z}$.  The quantization axis is taken along the ODT beam polarization, \textit{i.e.} along the vertical axis. The wavelength and polarization of the ODT beam are chosen such that the transitions $m=0 \rightarrow$ $m'=\pm 1$ of the intercombination are in the so-called magic wavelength configuration \cite{Ido2003}. More precisely, the ODT-induced light-shift of those Zeeman substates is identical. Hence, once the LRI beams are on, the presence of the ODT does not lead to extra spatially dependent radiation pressure forces that might compete with the shadow force under investigation. The polarizations of the LRI lasers lie in the $xy$ plane (see Fig. \ref{fig:setup}) and thus, address the transitions $m=0 \rightarrow$ $m'=\pm 1$ as expected. In addition, the earth magnetic field is compensated below the milliGauss level, so that the Zeeman frequency shift of the atomic states can be disregarded \cite{pandey2016}.
	
	The ODT beam is focused along its vertical transverse dimension to realize the strong aspect ratio between the weak confinement in the horizontal plane and the stronger confinement in the vertical dimension. We get the beam waists of $w_Y=138\,\mu$m and  $w_z=14\,\mu$m along the axes $\hat{Y}$ and $\hat{z}$ respectively, whereas the Rayleigh length -along $\hat{X}$- 
	is $360\,\mu$m (see  Fig. \ref{fig:setup}). The ODT power is $0.95\,$W, leading to a trap depth of about $25\,\mu$K and trapping frequencies of
	 $11.5\,$Hz, $30\,$Hz and $300\,$Hz along the $\hat{X}$, $\hat{Y}$, and $\hat{z}$ axes respectively. We load typically $3\times 10^5$ atoms at a temperature of $1\,\mu$K, leading to cloud sizes, at equilibrium and without the LRI beams, around
	  $L_X=140$ $\mu$m, $L_Y=50$ $\mu$m  and $L_z=5$ $\mu$m. The temperature remains almost constant in the $xy$ plane. In addition, a dimple trap beam at $852$ nm propagating along $\hat{z}$ allows for further trapping in the horizontal plane. The dimple beam has a waist of 80 $\mu$m and power 180 mW at the level of the cloud. The role of the dimple trap consists in reducing the size of the atomic cloud, helping to reach the stationary regime of artificial gravity experiment in a shorter time. The dimple beam is switched off when the LRI beams are turned on. 
	
	The intensity of each LRI beam is balanced independently using half wave plates and polarizing beamsplitters. Intensity balance is realized when the cloud's center stays at a fixed position all along the experiment.
	
	The experiment is done by varying $s_0$ in the range of 0.2 to 2, and for each $s_0$, the duration of the LRI beams (before imaging) is varied up to 100 ms.

	\subsection{Imaging scheme}
	
	\begin{figure}[t!]
		\flushleft
		\includegraphics[width=0.48\textwidth]{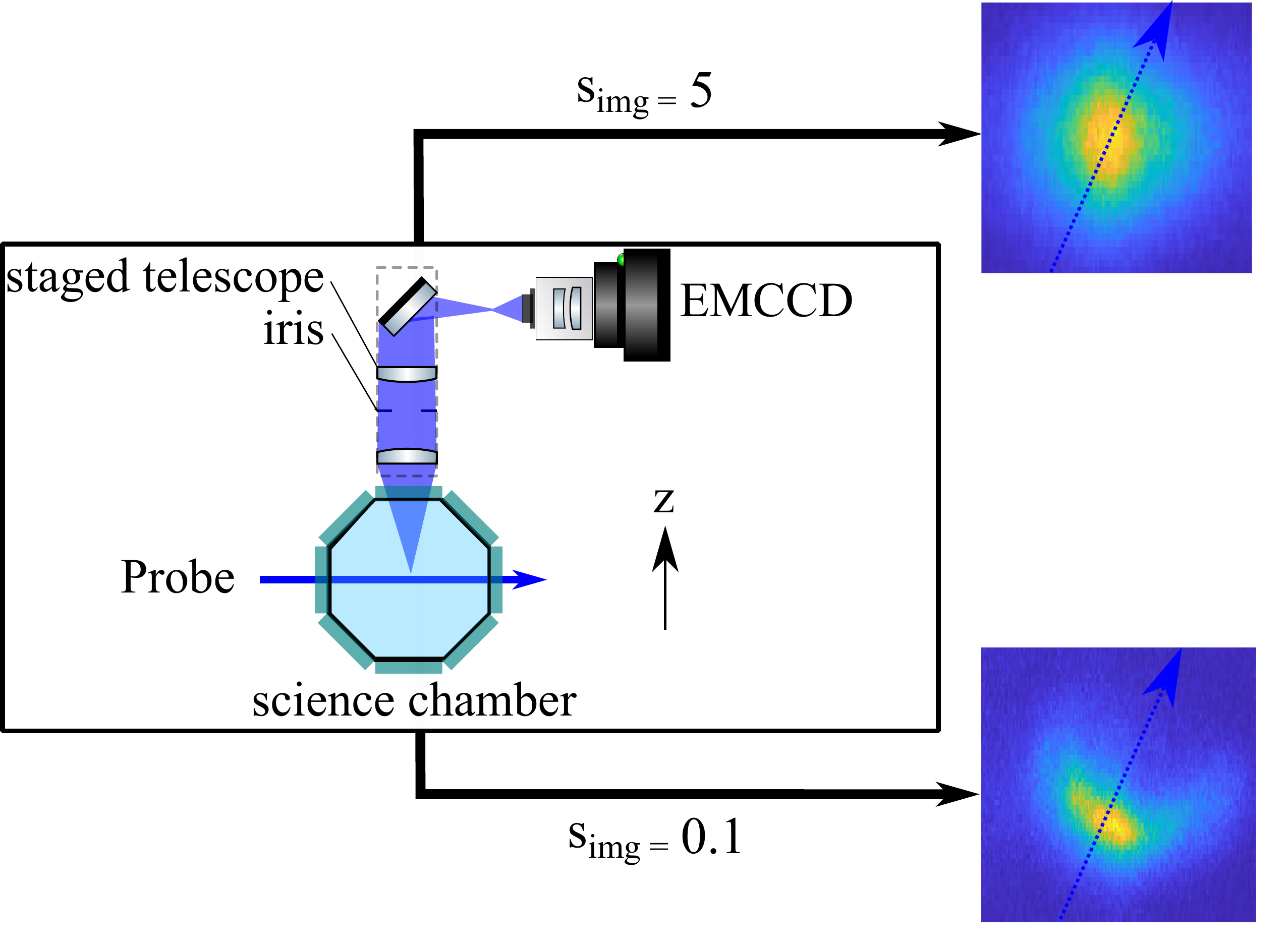}
		\caption{Schematic of the imaging system and fluorescence imaging . The blue dotted arrow shows the direction of propagation of the resonant $461\,$nm probe. Right panels: The false color fluorescence images are obtained for saturating (upper) and non-saturating (lower) probes. The counts of the non-saturated fluorescence image is multiplied by two for readability. $s_{\rm img}$ corresponds to the saturation parameter of the probe.}
		\label{fig:imaging}
	\end{figure}
	
	The analysis of the atomic cloud is performed thanks to a fluorescence imaging system having its optical axis along the vertical axis (see Fig. \ref{fig:imaging}). The probe laser is tuned on resonance with the dipole allowed transition at $461\,$nm, for optimal signal-to-noise ratio. The probe beam makes a $30^{\circ}$ angle with the $\hat{x}$ axis. Due to the low optical depth of the cloud along the $\hat{z}$ direction, the integrated fluorescence signal is proportional to the atoms number. The coefficient of proportionality is extracted thanks to a preliminary joint absorption and fluorescence imaging measurement. The atomic cloud is fully characterized using three fluorescence images. A first one, at high saturation intensity, is taken after turning off the LRI beams. This image allows us to extract the cloud size in the horizontal plane. Since the saturation is high, the absorption of the probe is weak, which gives a precise (\emph{i.e.} a relative statistical error below $10\%$) estimation of the atoms number (see a sample image in right upper panel in Fig. \ref{fig:imaging}). A second image, at low saturation, is also taken after extinction of the LRI beams. In this case, absorption of the probe is clearly visible (see a sample image in right lower panel in Fig. \ref{fig:imaging}), allowing for optical depth measurements. With those two images, we perform a full characterization of the sizes and optical depths of the atomic cloud. 
	
	Importantly, the measured optical depths correspond to the $461\,$nm transition and thus, need to be transposed to the $689\,$nm transition of interest. Since, the latter transition is narrower, Doppler broadening shall be considered \cite{Yang2015}. To do so, we measured the temperature of the cloud using a third image at high saturation taken at long time (typically $300\,$ms) after turning off the LRI beams. In this case, the cloud has reached thermalization in the ODT, so the horizontal temperatures can be extracted from the cloud sizes $L_i^{\rm th}$ ($i=X,Y$), and the trapping frequencies. Moreover, comparing the cloud size before and after thermalization time gives access to 
	\begin{equation}
	\text{CF}_i(t_\text{exp})=\frac{L_i^{\rm th}}{L_i(t_\text{exp})}.
	\end{equation}
	the compression factor experienced by the cloud due to the 2D gravity effective interaction. Here, $t_\text{exp}$ corresponds to the duration of the 2D gravity experiment.
	
	\subsection{Atoms losses}
	\label{sec:evaporation}
	
\begin{figure}[t!]
		\flushleft
		\includegraphics[width=0.44\textwidth]{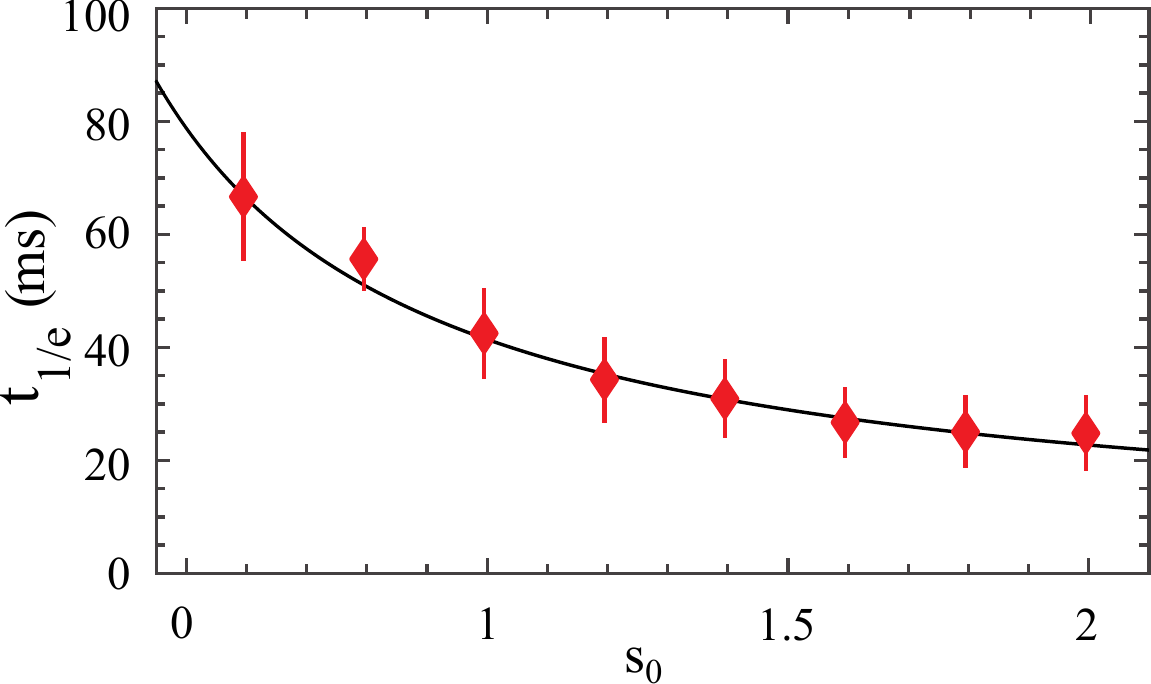}
		\caption{(left) Atom $1/e$-lifetime in the presence of 2D gravity beams for $\bar{\delta}=-3$ and $N_0\simeq 3\times 10^5$ from  data exponential fit (red diamond) and analytical expression of Eq. \eqref{eq:t_1/e} (solid line).
			\label{fig:lifetime}}
	\end{figure}

	In Fig. \ref{fig:lifetime}, we plot the lifetime of the cold atomic cloud in the ODT as a function of the saturation parameter $s_0$ of the long-range force laser beams. Without the LRI beams, the lifetime is above 20 s. The strong reduction of the cloud lifetime, when the LRI beams are on, is due to atoms spilling out of the trap along the uncooled vertical direction. When compression occurs in the $xy$ plane due to long-range 1D+1D forces, the temperature is expected to remain approximately constant in-plane because of the optical molasses, but to increase along the vertical direction $\hat{z}$. As atoms gain mechanical energy, they overcome inevitably the trap depth $U_0$ at some point and are removed from the system. 
	We provide now a simple model quantifying this effect. We consider the escape process as a single particle problem and discard the spatial distribution of atoms in the trap. At the beginning of the gravitational experiment, atoms are thermalized, and their temperature is much less than the trap temperature $U_0/k_B$. Hence, we take the initial atom energy to be zero and consider its increase due to spontaneous emission. 
	We assume each scattering event increases the kinetic energy of an atom in the vertical direction by $E_r$, of the order of the photon recoil energy. Since  $n=\frac{U_0}{E_r}\simeq 110$ is large, each atom will undergo many scattering events before leaving the cloud, and this approximation should be reasonable.
	The effective scattering rate is given by (see for instance \cite{Steck2019})
	\begin{equation}
	\xi=\frac{12}{5} \Gamma \frac{s_0}{1+4\bar{\delta}^2}.
	\label{eq:Gamma_eff}
	\end{equation}
	The prefactor comes from the radiation pattern and number of LRI beams. 
	Since the photon scattering rate $\xi$  is constant, the number $N_S(t)$ of scattering events follows a Poissonian distribution with parameter $\xi t$, and the energy of an atom is $E(t)=E_r N_S(t)$. At time $t$, the fraction of atoms remaining in the trap is then given by $\mathbb{P}(N_S(t) < n)=F(\xi t,n)$, where $F(\lambda,n)$ is the cumulative distribution function of a Poisson variable with parameter $\lambda$. The characteristic $1/e$-lifetime corresponding to $N(t_{1/e})=N(t=0)/e$ can be obtained using the Stirling's formula for the incomplete Gamma function; one gets
	
	\begin{equation}
	t_{1/e}\approx \frac{n}{\xi}
	\label{eq:t_1/e}.
	\end{equation}

	Fig. \ref{fig:lifetime} gives a very good agreement between the experiment and the model of Eq. \eqref{eq:t_1/e} when evaluating the characteristic $1/e$-lifetime of atoms within the trap.
	This quantitative evidence suggests that indeed single atom heating and spilling along the vertical direction is at the origin of the atom losses.
	
	\subsection{Experimental results}\label{sec:exp_results}
	
	\begin{figure}[t!]
		\flushleft
		\includegraphics[width=0.48\textwidth]{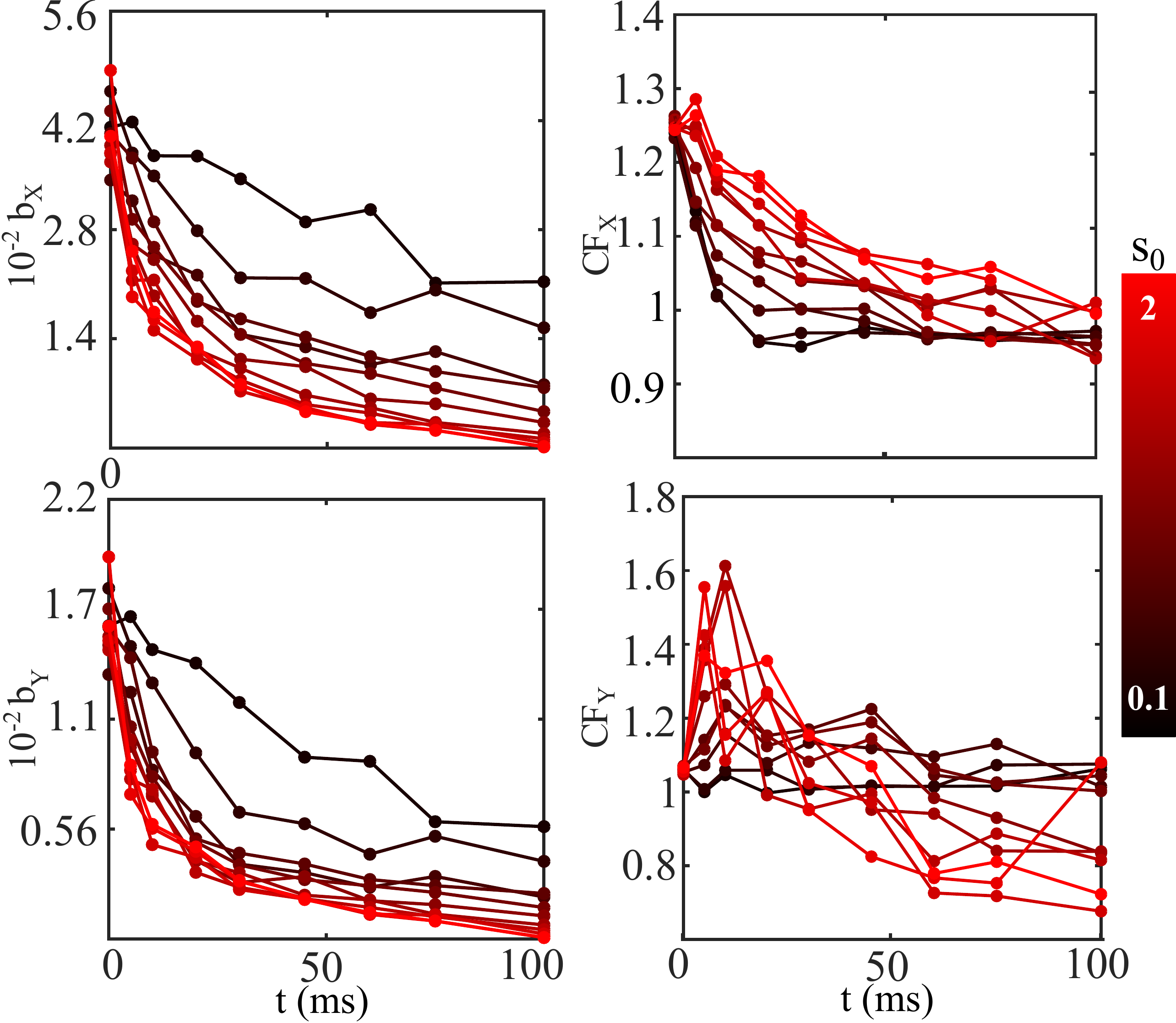}
		\caption{Optical depths at detuning  $\bar{\delta}=-3$ (left column) and compression factors (right column) along $\hat{X}$ and $\hat{Y}$ for a temperature $\theta\approx 1\,\mu$K, and an initial atoms number $N_0\approx 3\times 10^5$.}
		\label{fig:compression_experiment}
	\end{figure}
	
	The temporal evolution of the optical depth and compression factor along $\hat{X}$ and $\hat{Y}$ (the ODT proper axes) for various saturation parameters are given in Fig. \ref{fig:compression_experiment}. According to the 2D model (see Sec. \ref{sec:sgb} and Fig. \ref{fig:phase_diagram_2D}), the system should be in the collapsed phase, at least for the large values of the saturation parameter. However, we did not observe the expected signatures of the collapsed phase which are an increasing of the compression factor and a saturation of the optical depth to a value around one. Indeed, after a short time of $5-10\,$ms 
	we observe a compression of the cloud of about $60\%$ in the $\hat{Y}$ direction and a more moderate compression of 
	about $30\%$ in the $\hat{X}$ direction. As expected, those compression factors are higher for larger saturation parameter, but the compression factor rapidly falls to value close to one. The optical depth (left column) is initially rather large, which might explain the initial moderate compression. However, we observe a monotonous decrease of the optical depth without any sign of saturation. We observe that the decrease of the optical depth is more pronounced at large saturation in agreement with a larger atomic loss rate (See Fig. \ref{fig:lifetime}). The compression factors larger than one at $t=0$ originate from the vertical dimple trap beam as shown by the green arrow on Fig. \ref{fig:setup}. This beam is turned off at $t=0$.
	
	\section{Three-dimensional Model}\label{sec:model}
	
	In the previous section, the experimental results show a moderate compression of the atomic cloud, but no signature of a collapsed phase as predicted by the 2D model of Sec. \ref{sec:2Dsgs}. In order to  be closer to the experimental reality, we generalize  
	this model in 3D, including the finite thickness of the cloud in the vertical direction, and the Coulomb-like repulsion induced by multiple scattering.
	
	\subsection{Description of the model}
	\label{Sec:Des_model}
	For a 3D model, the  particle dynamics is now driven by three forces depending on the particle position,  $\mathbf{F}_{\rm tot}=\mathbf{F}_\text{2D}+\mathbf{F}_\text{T}+\mathbf{F}_\text{M}$ and by $\mathbf{F}_\text{d}$,  the friction force associated with  Doppler cooling.
	
	$\mathbf{F}_\text{2D}$ is the attractive force due to the LRI beams, $\mathbf{F}_\text{T}$ the harmonic trapping force, and $\mathbf{F}_\text{M}$  is the repulsive force coming from multiple scattering, which cannot be discarded in a three dimensional geometry. The expression of the attractive force is: 
	\begin{align}
		\mathbf{F}_{\text{2D}}(\mathbf{r}) &= -C' \left(\begin{array}{c}  \int dx' \text{sgn} (x-x') \rho(x',y,z)\\  \int dy' \text{sgn} (y-y') \rho(x,y',z)\\ 0  \end{array}\right) , \label{eq:forceG}
	\end{align}
	where 
	\begin{equation}
	C'=\frac{\hbar k\Gamma}{2}s_0\frac{N\sigma_0}{(1+4\bar{\delta}^2)^2}.
	\end{equation}
	Here, the LRI beams lead to the same force than derived in 2D.
	
	To mimick the experiment, we consider an anisotropic trap, with its principal axes along the unit vectors $\hat{X}$, $\hat{Y}$,$
	\hat{z}$ ($\hat{X},\hat{Y}$ are not aligned with the LRI beams, see Fig.\ref{fig:setup}). The associated force at a position 
	$\mathbf{R}=X\hat{X}+Y\hat{Y}+z\hat{z}$ is: 
	\begin{align}
		\mathbf{F}_\text{T} (\mathbf{R})&= - m\left[ \omega_X^2X \hat{X} +\omega_Y^2Y \hat{Y}+ \omega_z^2 z \hat{z} \right]. \label{eq:forceT}	
	\end{align}
	The friction force $\mathbf{F}_\text{d}$ has the same expression as in the 2D model. Finally, the repulsive force coming from multiple scattering is given by \cite{Sesko1991}
	
	\begin{align}
		\mathbf{F}_\text{M}(\mathbf{r})& =D  \int d^3r' \frac{\mathbf{r}-\mathbf{r'}}{|\mathbf{r}-\mathbf{r'}|^3 }\rho(x',y',z') , \label{eq:forceM}	
	\end{align}
	where 
	\begin{equation}
	D=\frac{\hbar k\Gamma}{2}s_0\frac{N\sigma_0\sigma_R}{\pi(1+4\bar{\delta}^2)}.
	\end{equation}
	and $\sigma_R$ is the re-absorption cross-section of scattered photons. We assumes an isotropic fluorescence pattern and multiple scattering limited to a single re-absorption event. The latter is well justified in the low optical depth regime. The former overestimates the multiple scattering contribution with respect to the experiment where the LRI beam polarization are in the horizontal plane, leading to a radiation pattern more pronounced (by a factor of two) along the vertical direction.
	
	To make the comparison between the strength of the attractive and repulsive forces easier, we write the equations in a slightly different way than in Sec.\ref{sec:2Dsgs}.
	By introducing  the length scale $L$ 
	\begin{equation}
	L^3=\frac{2 \hbar k_L \Gamma \sigma_R}{m \omega^2},
	\end{equation}
	and the timescale $\tau$	
	\begin{equation}
	\tau= \left(\frac{\eta}{\omega^2}\right),
	\end{equation}
	where $\omega$ is a characteristic trap frequency in the $xy$ plane (the actual frequency is not the same along the $X$ and $Y$ axes).
	Then the Smoluchowski equation can be expressed in a dimensionless form
	\begin{equation}
	\frac{\partial \rho}{\partial t} = \bm{\nabla} \cdot \left[ -\rho \mathbf{F}_{\rm tot} + T \bm{\nabla} \rho \right], \label{eq:smol}
	\end{equation}	
	where the dimensionless temperature $T$ is now given by
	\begin{equation}
	T=\frac{k_B\theta}{m\omega^2 L^2}= \frac{k_B \theta}{(2\sqrt{m}\omega  \hbar k_L  \Gamma \sigma_R)^{2/3}},
	\end{equation}
	and the dimensionless forces can be written as
	\begin{eqnarray}
	\mathbf{F}_{\text{2D}}(\mathbf{r}) &=& -\frac{\gamma c}{4} \left(\begin{array}{c}  \int dx' \text{sgn} (x-x') \rho(x',y,z)\\  \int dy' \text{sgn} (y-y') \rho(x,y',z) \end{array}\right) , \label{eq:force1} \nonumber \\
	\mathbf{F}_\text{M}(\mathbf{r})& =&\frac{c}{4 \pi}  \int d^3r' \frac{\mathbf{r}-\mathbf{r'}}{|\mathbf{r}-\mathbf{r'}|^3 }\rho(x',y',z') , \label{eq:force2} \\
	\mathbf{F}_\text{T} (\mathbf{R})&=& - \left[ \left(\frac{\omega_X}{\omega}\right)^2X \hat{X} +\left(\frac{\omega_Y}{\omega}\right)^2Y \hat{Y}+ \left(\frac{\omega_z}{\omega}\right)^2 z \hat{z} \right]. \nonumber	
	\label{eq:force3} 
	\end{eqnarray}
	The parameters $c$ and $\gamma$ are given by
	\begin{equation}
	c = \frac{N s_0}{(1+4\bar{\delta}^2)^2}~{\rm and}~\gamma = \frac{\sigma_0}{\sigma_R},
	\end{equation}
	Note that, in the low saturation regime, the scattering of re-emission is elastic (no change of photon frequency) 
	meaning 
	that $\gamma=1$, reaching its maximal value. If saturation of the transition occurs, 
	the scattering becomes inelastic and part of the fluorescence spectrum is brought at the transition 
	resonance \cite{Mollow1969}, increasing $\sigma_R$. 
	Following \cite{cohen1998atom}, we estimate $\gamma \approx 0.98$ for a saturation parameter $s_0 = 1$ per beam.
	 This is the value we have used in the simulations.
	
	\begin{figure}[t!]
		\includegraphics[width=0.42\textwidth]{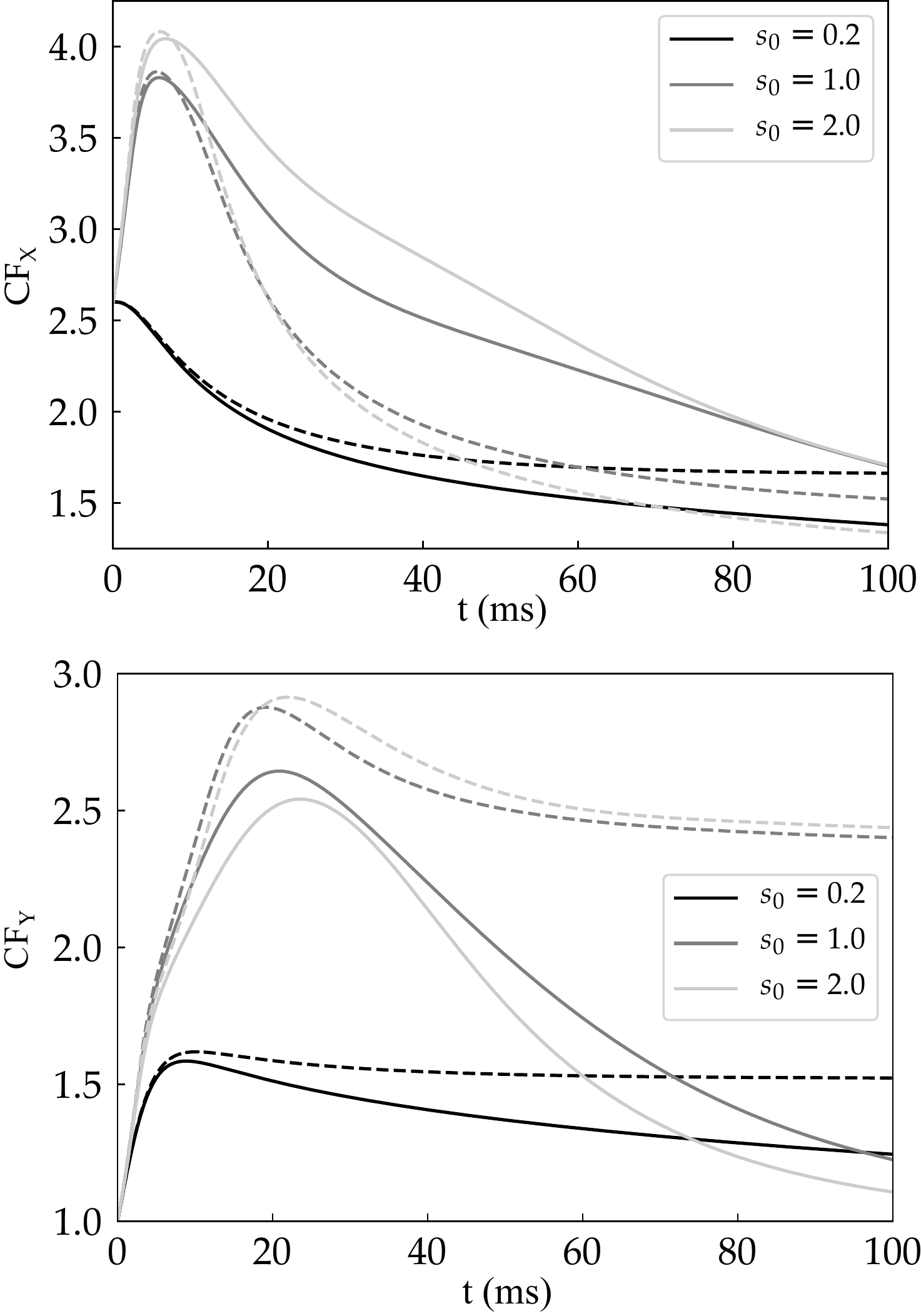}\\
		\caption{\label{fig:simulations} Time evolution of the compression factors along the two directions $\hat{X}$ (up) and $\hat{Y}$ (bottom),
			obtained by direct simulation of the two 3D models:   
			The dashed curves correspond to the model with a constant number of atoms and the full curves to the model including effective atoms losses. The angle between the axes $\hat{x}$ and $\hat{X}$ is set to $30^{\circ}$ (to  be compared to $23^{\circ}$ in the experiment). The corresponding dimensionless temperature of the 2D model for $s_0=1$ is $\sim 0.06$, \textit{i.e.} in the predicted collapsed phase according to Sec.\ref{sec:2Dsgs}.}
	\end{figure}

	\subsection{Numerical simulations}
	\label{sec:simulations}
	
	We have performed two types of simulations of Eqs. \eqref{eq:smol},\eqref{eq:force2}: i) with a fixed number
	 of atoms; ii) including atoms losses in an effective manner: interaction forces then have an exponentially
	  decreasing strength $c(t)=c_0e^{-t/t_{1/e}}$; we have taken $t_{1/e}$ as given by the curve 
	  in Eq.~\eqref{eq:t_1/e}.
	
	A proper numerical integration of the Smoluchowski-Poisson equation, given by Eq. \eqref{eq:smol}, is 
	a challenging task. The repulsive force is computed via a standard Poisson solver for the Coulomb-like 
	potential. For the attractive force, we note that, although it does not globally derive from a potential, 
	the $x$ and $y$ components of the force taken separately do. We then use 
	the finite-volume method presented in \cite{Carrillo2015}, which is well suited to obtain solutions at 
	long times for such potential forces, and we couple it with a splitting procedure: we first compute a 
	time step with only the $x$ component of the attractive force, then a time step with only its $y$-component, 
	and repeat. A further numerical difficulty is related to the spatial scale difference between the $xy$ plane
	 and the transverse $z$ direction.
	
	Because the compression observed in experiment is a transient phenomenon, we focus our study of the $3D$ models 
	on the time evolution in order to compare more efficiently the numerical results to our experiment. 
	We first consider
	the time evolution of the compression factors for the two 3D models (see Fig. \ref{fig:simulations}).
	The values of the simulation parameters  are chosen in agreement with the experimental ones. 
	The dashed curves correspond to a constant atoms number, and the plain curves include an exponential decay 
	of the atoms number, with $t_{1/e}$ chosen as in Eq. \eqref{eq:t_1/e}, for three different values of 
	the saturation parameter $s_0=0.2,\,1,\,2$. The initial number of atoms is $N_0=3\times 10^5$,  
	the detuning is $\bar{\delta}=-3$, and temperature is $\theta=1\,\mu$K. The planar anisotropy of 
	the trap is set by the ratio $\omega_z/\sqrt{\omega_X\omega_Y}=15$, and the ratio of the trap 
	frequencies in the $xy$ plane is $\omega_Y/\omega_X = 2.6$. Since the initial density distribution
	 $\rho(t=0)$ is chosen as an isotropic Gaussian in the $xy$ plane, with thermal width corresponding 
	 to $\omega_X$, the initial value of  the compression factor along the 
	 $X$ axis is $\text{CF}_X(t=0)= 2.6$.
	
		\begin{figure}[t!]
		\includegraphics[width=0.45\textwidth]{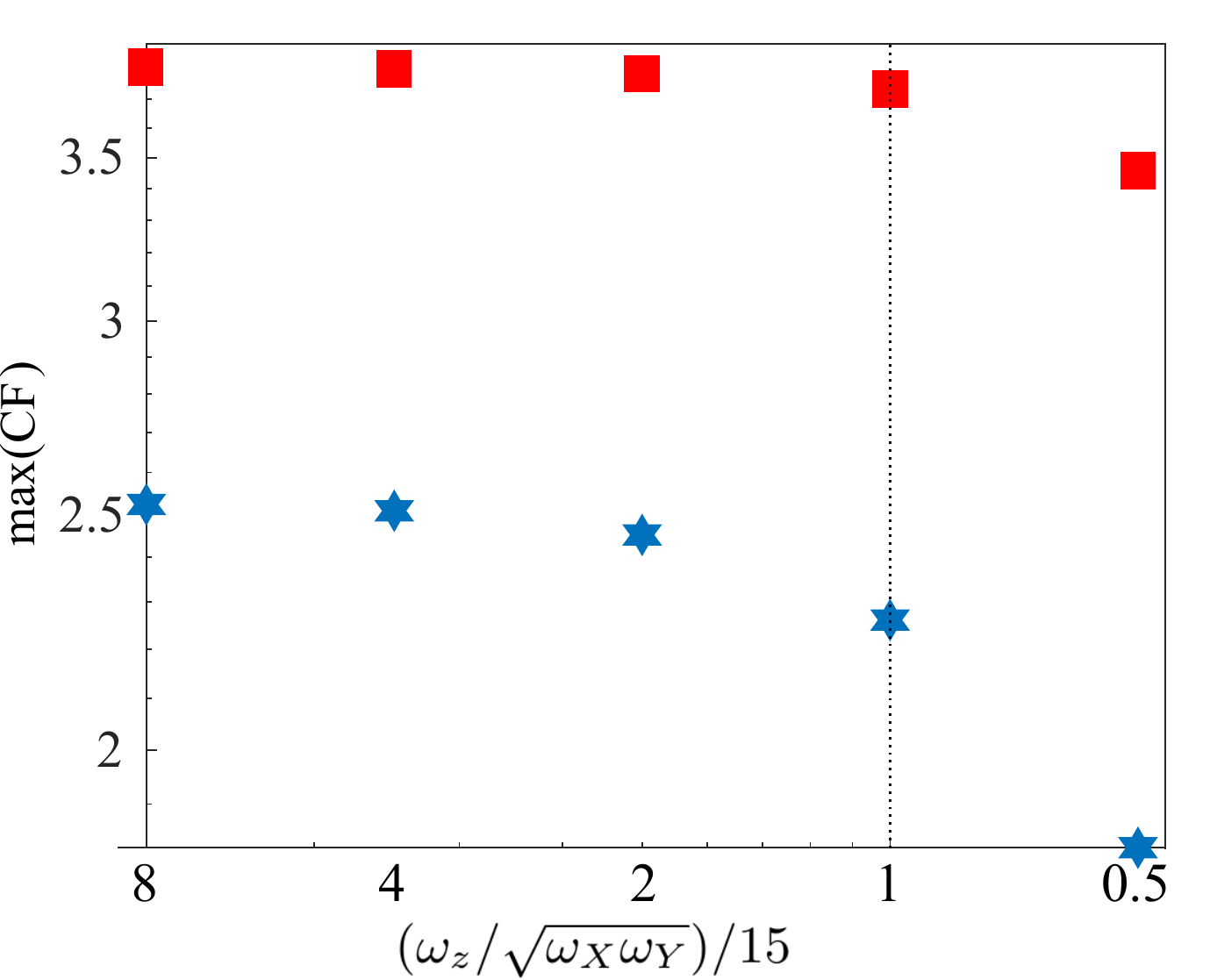}
		\caption{\label{fig:maxCF} Maximum transient compression factors along the two trap directions 
			(red squares) $\hat{X}$ and (blue stars) $\hat{Y}$ as a function of the normalized trap ratio for $s_0=1$. $\omega_z/\sqrt{\omega_X\omega_Y}=15$ corresponds to the experimental value, indicated by the vertical dashed line. 
			Simulation results correspond to the model where the number of atoms is constant.
			The shift between the compression factors  between the $X$-direction
			and the $Y$-direction are due to the trap anisotropy in the horizontal plane.}
	\end{figure}
	
	Comparing simulation results on Fig. \ref{fig:simulations} to the experimental ones 
	in Fig. \ref{fig:compression_experiment}, we note a  qualitative agreement: when  the  saturation parameter is 
	not too small, the compression factor increases at short time,
	 reaches a maximum for $t \sim 5-20\,$ms, and then decreases.  
	However, simulations predict larger compression factors than those observed. 
	The discrepancy between the model and the experiment may come from the assumptions made in the $3D$ model: 
	the long range of the shadow force is associated with the linearization of the Lambert law, 
	leading to a stronger compression force. At longer time ($>20$ ms), atoms losses drive the system 
	towards a trivial stationary state, without compression. 
	
	When the number of atoms is kept constant, (dashed curves in Fig. \ref{fig:simulations}), 
	we observe the same behavior at short time, \textit{i.e.} when the atoms losses are not significant. 
	At long time ($\gtrsim 20\,$ms), the system settles in this case in a stationary compressed state. 
	Comparing with the $2D$ model, we conclude that the 3D effects
	play a significant role to explain the relatively small observed values of the compression.

	We now investigate the role of the trap aspect ratio 
	$\omega_z/\sqrt{\omega_X\omega_Y}$; to do so, we keep constant the initial optical depth in the plane $z=0$, 
	which amounts to keep constant the rescaled temperature $T$ in the associated $2D$ model,
	 well inside the "collapse region". Fig. \ref{fig:maxCF} displays the compression factor versus the
	  trap aspect ratio. 
	We expect that the behavior of the system becomes closer to the prediction of the $2D$ model when this aspect ratio tends to infinity.  However, the 3D model predicts that the size of the system saturates to a finite value, 
	in contrast to the infinite compression predicted by the collapsed phase of the $2D$ model.

\begin{center}
		\begin{figure}[t!]
			\includegraphics[width=0.95\textwidth]{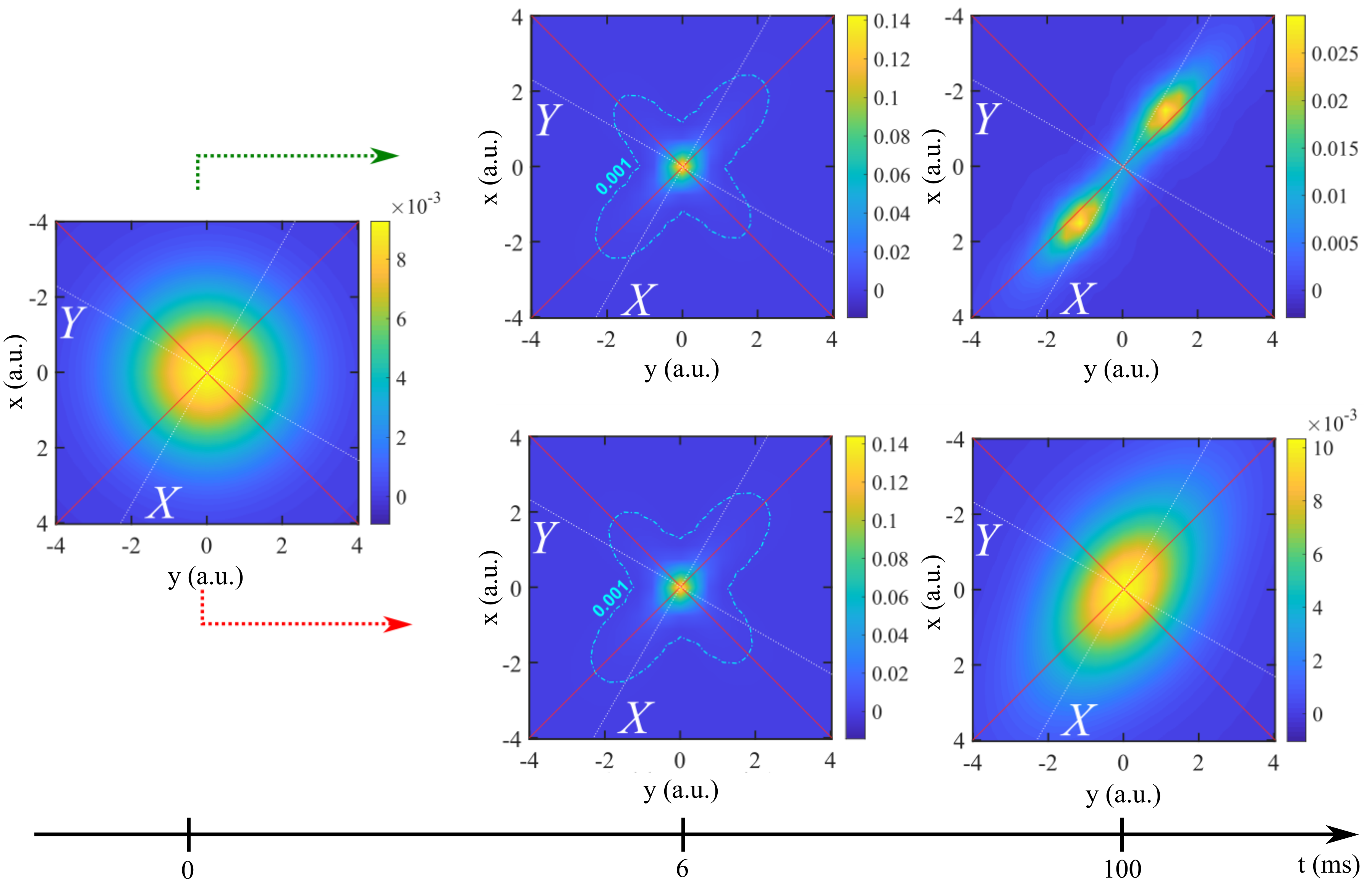}
			\caption{\label{fig:snapshots} Simulated spatial distributions of atoms in the plane $z=0$ at $t=0$ 
				(left), at $t=6$ ms (middle) corresponding to the maximum of the compression factor along $\hat{X}$,
				 and at $t=100$ ms (right).  $s_0=1$. Upper panels (top green dotted arrow) correspond to the model 
				 with a constant number of atoms and lower panels (bottom red dotted arrow) to the model with an 
				 exponential decrease of the atom number. We also draw  the principal axes 
				 of the trap (white dashed) and the first bisector angle between LRI lasers (red continuous). 
				 Contour lines, for intermediate time, highlight the star-shape of the cloud 
				 as predicted by the 2D model (see \citep{Barre2014}). a.u. stands for arbitrary units.} 
		\end{figure}
	\end{center}

	To illustrate the time evolution, Fig. \ref{fig:snapshots} shows several snapshots of the atomic spatial 
	distribution in the $z=0$ plane at three different times: $t=0$,  $t=t_m=6$ ms corresponding 
	to the maximum compression along $\hat{X}$ and at the final time $t = 100$ ms of the simulation. 
	We first focus on the upper row which corresponds to the case of constant atom number. 
	We observe a fast compression of the cloud as also indicated by the compression 
	factor in Fig. \ref{fig:maxCF}; the cloud at $t=t_m$ has a long-range star-shape similar to the one observed in the $2D$ 
	simulations in \cite{Barre2014}, 
	 cyan dash-dotted contour line helps to visualize this shape.  
	At $t=100$ms, the system has essentially reached a stationary state, but it displays unexpected spatial patterns: 
	the cloud has split into two parts along one bisector of the LRI lasers, where the attractive long-range force 
	is the weakest. This rich dynamic results from a competition between the geometry-driven attractive and 
	repulsive long-range interaction. A detailed study of this phenomenon is beyond the scope of paper. 
	We focus now on the lower row of Fig. \ref{fig:snapshots}, corresponding to the case with atoms losses. 
	As already discussed, the effect of interactions becomes negligible at large times, a decompression occurs
	 and the cloud's shape corresponds to the expected shape in the harmonic trap at thermal equilibrium.
	
	\onecolumngrid
	\twocolumngrid

	\section{Conclusion}\label{sec:conclusion}
	
	We have studied the interaction of a quasi-two-dimensional ultra-cold atom cloud with
	 two orthogonal quasi-resonant counter-propagating pairs of lasers.
	  For low optical depth, each pair of laser mimics one-dimensional artificial 
	  gravity-like long-range force. We have reached experimentally the regime where 
	  a two-dimensional 
	  analysis predicts a collapse of the cloud, but we have observed only a 
	  moderate compression. 
	  To understand this discrepancy, we have introduced a three-dimensional model
	   which provides a more 
	  realistic description of the cold atom cloud, and in particular includes the repulsive long-range force 
	  coming from photons reabsorption: our results show that although repulsive long-range force is
	  	 partly suppressed by the pancake-shape geometry as expected, it is non negligible when 
	  	 the compression occurs. If we include atoms losses along the uncooled vertical dimension, 
  	 the model is in qualitative agreement with the experiment. We conjecture some of the remaining 
  	 discrepancies may be due to the low optical depth approximation which is implemented in the theoretical 
  	 approaches. 
  	 Nevertheless, the satisfactory behavior of our three-dimensional numerical model makes it a useful 
  	 tool to investigate new regimes of the atomic cloud and design new experiments: in particular, 
  	 it predicts a rich spatio-temporal behavior related to the peculiarities of the interplay between 
  	 shadow attractive and repulsive scattering effects. Its experimental investigation would require 
  	 increasing the atom lifetime in the trap; this could be obtained for example by increasing the trap depth
  	  or by adding two extra counter-propagating vertical cooling laser beams. 
  	  The presence of extra vertical beams will increase 
  	  	the scattering rate in the horizontal
  	  	 plane,
  	  	 but we still expect that the attractive shadow interaction 
  	  	 dominates for large aspect ratio.
	
%

\end{document}